\documentclass[epj]{svjour}
\usepackage{graphics, color}
\usepackage{bm}       
\usepackage{amsmath}  
\usepackage{amssymb}
\usepackage{mathtools}

\usepackage{footnote}
\newcommand{\bra}{\langle}
\newcommand{\ket}{\rangle}
\newcommand\CO{{\cal O}} 

\begin{document}
\title{First principles electromagnetic responses in medium-mass nuclei}
\subtitle{Recent progress from coupled-cluster theory}
\author{Johannes Simonis\inst{1} \and Sonia Bacca\inst{1} \and Gaute Hagen\inst{2,3}
}                     
\offprints{}          
\institute{Institut f\"{u}r Kernphysik and PRISMA Cluster of Excellence, Johannes Gutenberg-Universit\"{a}t, Mainz, DE-55128, Germany \and Physics Division, Oak Ridge National Laboratory, Oak
  Ridge, Tennessee 37831, USA \and Department of Physics and Astronomy, University of Tennessee, Knoxville, TN 37996, USA} 
\date{Received: date / Revised version: date}
%
\abstract{
  We review the recent progress made in the computation of
  electromagnetic response functions in light and medium-mass nuclei
  using coupled-cluster theory. We show how a many-body formulation of
  the Lorentz integral transform method allows to calculate the
  photoabsorption cross sections of $^{16, 22}$O and $^{40}$Ca. Then,
  we discuss  electromagnetic sum rules, with particular
  emphasis on the electric dipole polarizability, $\alpha_D$. By
  including triples corrections in coupled-cluster theory, we revisit
  $^{48}$Ca, for which, beside the electric dipole
  polarizability, we had previously investigated the neutron and
  proton radii, as well as the size of the neutron-skin
  thickness~\cite{Hagen2016}. We show that correlations among these
  observables still hold, albeit a  better agreement with
  experiment is obtained for $\alpha_D$ and the prediction of a small neutron-skin
  thickness is further corroborated.
} 
\maketitle
\section{Introduction}
\label{intro}

Electromagnetic probes are invaluable tools to study the nature of
composite quantum-mechanical systems, such as nuclei. Due to the small
value of the electromagnetic coupling constant $\alpha$, one can
cleanly relate measured cross sections to properties of the composite
system via perturbation theory, leading to a more complete
understanding of the internal dynamics.  Indeed, electromagnetic
probes have historically enabled important discoveries regarding the
nucleus and the strong dynamics governing its multifaceted properties.
Most notably, the study of photonuclear reactions lead to the
discovery of giant dipole resonances and to their interpretation in
terms of collective modes~\cite{GoT48,steinwedel1950}. A complete body
of data has been collected over the past decades for stable nuclei,
and some selected studies were performed even on unstable nuclei,
leading, e.g., to the discovery of pygmy resonances, see, e.g.,
Ref.~\cite{Bracco2019} and references therein.  But where do we stand
with the theory today?

One of the goals of modern nuclear theory is to be able to explain
nuclear phenomena starting from protons and neutrons as degrees of
freedom and by connecting their interactions to quantum chromodynamics
via the use of chiral effective field
theories~\cite{Weinberg90,Epelbaum09,Machleidt11,Epelbaum12} or with
other more traditional potentials~\cite{Wiringa1995}. This
research path is called ``ab initio approach'' in low-energy nuclear
physics, see,
e.g.,~Refs.~\cite{Leidemann12,Bacca:2014tla,Hebeler2015}.  What is
meant by that is that, for a given interaction Hamiltonian, the
quantum-mechanical problem of protons and neutrons interacting with
each other is either solved exactly, or within controlled
approximations~\cite{morten}.

\begin{figure}[ht]
\centering
 \includegraphics*[width=9cm]{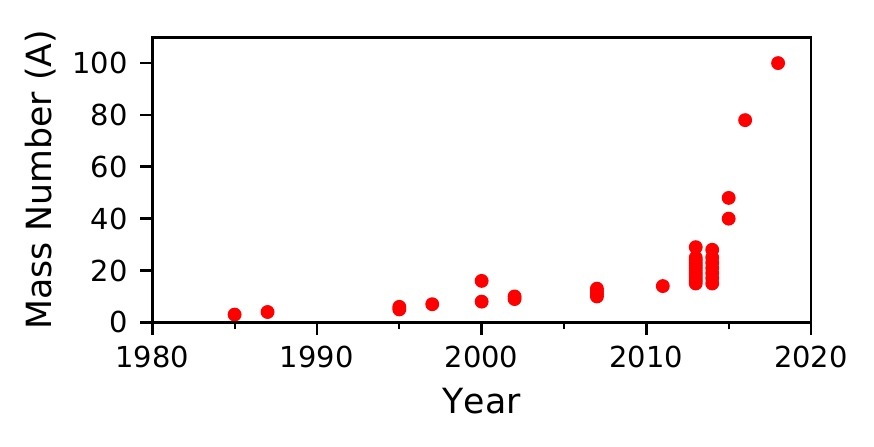}
 \caption{\label{fig_comp} Trend for realistic first principle
   calculations for the $A$-body nuclear problem as a function of $A$.
   Figure adapted and updated from Ref.~\cite{Hagen2016}.}
\end{figure}

In the last years, ab initio computations in nuclear physics have
advanced tremendously. While until one or two decades ago it was
possible to only deal with very few nucleons, today one can reach even
mass number $A\sim 100$~\cite{hagen2016b,simonis2017,Sn,gysbers2019}
and above.  The resulting growth of first principle calculations is
well captured in Fig.~\ref{fig_comp}, where we display the trend for
ab initio calculations for the nuclear many-body problem as a function
of $A$.
In early decades the progress was linear in $A$, mostly due to
exponentially expensive algorithms implemented on machines with
exponentially growing computing power. More recently, newly developed
polynomial scaling algorithms, such as coupled-cluster
theory~\cite{hagen2014}, made it possible to exploit the exponential
growth in computing power, pushing the boundaries of first principle
calculations.  However, one has to note that this fast growing
progress regards mostly the computation of bulk properties such as
binding energies and radii. Where does this ab initio approach stand
with respect to the computation of electroweak reactions with nuclei?
Or, in other words, are we able to describe the above mentioned
collective modes in a microscopic way, starting from protons and
neutrons interacting through realistic forces?

Only recently we have paved the road to a qualitative and quantitative
description of electromagnetic reactions with nuclei from first
principles.  In this work, we will review the progress made in this
respect thanks to the introduction of a new technique obtained by
merging coupled-cluster theory~\cite{hagen2014} with the Lorentz
integral transform approach~\cite{efros1994,Efl07}, that led to a
novel computational tool to address
electromagnetic reactions and related observables in an ab initio
fashion~\cite{Bacca:2013dma}.

The paper is structured in the following way. We will first describe
the computational tools in Section~\ref{theory}, then, after a brief
explanation of the chiral interactions in Section~\ref{hamiltonians},
we will present results in Section~\ref{results}. Finally, we will
draw our conclusions and present an outlook for the future in
Section~\ref{conclusions}.

\section{Computational tools}
\label{theory}

\subsection{Coupled-cluster theory}
Coupled-cluster theory, as originally introduced by Coester and
K{\"u}mmel~\cite{coester1960}, is aimed at solving the Schr\"{o}dinger
equation for a many-body system. For a given Hamiltonian $\hat{H}$
describing the quantum system, one assumes that the correlated
many-body wave function can be written with an exponential ansatz
\begin{equation}|\Psi_0\rangle=\exp{(\hat{T})}|\Phi_0\rangle\,,
\label{anzatz}
\end{equation}
where $|\Phi_0\rangle$ is a Slater determinant of any kind. The
operator $\hat{T}$, typically expanded in $n$-particle--$n$-hole ($np$--$nh$) excitations
(or clusters) $\hat{T}=\hat{T}_1+\hat{T}_2+\dots+\hat{T}_A$, is responsible for introducing
correlations.  Using the formalism of second quantization the
correlation operators can be written as
\begin{eqnarray}
\nonumber
\hat{T}_1&=&\sum_{ia} t_{i}^a \hat{a}_a^\dag \hat{a}_i\,,\\
\nonumber
\hat{T}_2&=&\frac{1}{4}\sum_{ijab} t_{ij}^{ab} \hat{a}_a^\dag \hat{a}^\dag_b \hat{a}_j \hat{a}_i\,,\\
\hat{T}_3&=&\frac{1}{36}\sum_{ijkabc} t_{ijk}^{abc}\hat{a}_a^\dag \hat{a}^\dag_b \hat{a}^{\dag}_c \hat{a}_k \hat{a}_j \hat{a}_i\,,\\
\nonumber
\cdots & &\cdots
\label{top}
\end{eqnarray}
where indices $i, j, k,\dots $ label occupied single--particle (hole)
states in the reference Slater determinant, whereas the $a, b, c,
\dots $ indicate unoccupied (particle) states.
 
The many-body Schr\"{o}dinger equation for the ground state becomes
\begin{equation}
\overline{H}_N \vert \Phi_0 \rangle = E_0 \vert \Phi_0 \rangle \,,
\end{equation}
employing the similarity transformed 
Hamiltonian 
\begin{equation}\label{hbar}
 \overline{H}_N = \exp(-\hat{T})  \hat{H}_N \exp(\hat{T})\,,
\end{equation}
where $\hat{H}_N$ is normal ordered with respect to the reference Slater
determinant.  The amplitudes of the $\hat{T}$ operator, such as $t_i^a$ ,
$t_{ij}^{ab}$, $t_{ijk}^{abc}$, etc., are found by solving the
non-linear equations obtained by \begin{eqnarray} \nonumber
  0&=&\langle \Phi_i^a| {\overline H}_N| \Phi_0 \rangle\,,\\
  0&=& \langle \Phi_{ij}^{ab}| {\overline H}_N| \Phi_0 \rangle\,,\\
  \nonumber
  0&=& \langle \Phi_{ijk}^{abc}| {\overline H}_N| \Phi_0 \rangle\,,\\
  \nonumber &&\cdots \,.
\end{eqnarray}
Here, $\vert \Phi_{i}^a \rangle $, $\vert \Phi_{ij}^{ab} \rangle$ and
$\vert \Phi_{ijk}^{abc}\rangle$ are Slater determinants constructed as
$1p$--$1h$, $2p$--$2h$, $3p$--$3h$, $\dots$ excitations on top of the reference
state, respectively~\cite{hagen2014,shavittbartlett2009}.

Coupled-cluster theory is exact when the expansion of the $\hat{T}$ operator
is performed up to $Ap$--$Ah$
excitations. However, due to the exponential ansatz of
Eq.~(\ref{ansatz}), even when truncations are introduced, the result
is very close to the exact one. For closed (sub-) shell nuclei the
coupled-cluster method truncated at the $2p$--$2h$ level, at the so
called coupled-cluster singles and doubles level -- labeled with D in
this work--, captures about $90\%$ of the full correlation energy,
while adding triples excitations, about 99$\%$ of the correlation
energy is accounted for~\cite{bartlett2007,hagen2009b}. The advantage of
the method is that it scales polynomial with increasing system size
(by system size we mean a measure of the number of particles and the
size of the employed basis). For example, in the D approximation, the
algorithm scales as $n_o^2n_u^4$, where $n_o$ and $n_u$ are the number
of occupied and unoccupied orbitals, respectively.  Coupled-cluster
theory has been successfully applied to study properties of
closed (sub-) shell nuclei and their neighbors (see for example
Ref.~\cite{hagen2014} for a review, and references therein).

\subsection{Lorentz integral transform method}
While enormous progress has been made in first-principles computations
of ground-state properties of nuclei with increasing mass number $A$,
advances in the calculation of electromagnetic reactions with nuclei
have been slower, because of the additional challenges one has to
face.

Electromagnetic cross sections are typically related to the nuclear response function, defined as
\begin{equation} \label{response}
R(\omega)=
\sum_{n} | \langle \Psi_0
  |{\hat{\Theta}} 
   |\Psi_n\rangle|^2 
 \delta (E_{ n} -E_0 -\omega )\,.
\end{equation}
Here,  $|\Psi_0\rangle$ and $| \Psi_n \rangle$ denote ground and final state
wave functions of the nucleus, and  $E_0$ and
$E_{n}$ are their respective energies, with $\omega=E_n-E_0$. The operator ${\hat{\Theta}}$ 
is a generic electromagnetic operator.
One has to note that the $\sum_{ n}$ indicates both the sum over discrete states and an
integration over continuum Hamiltonian eigenstates. The calculation of the latter is the main bottleneck
in the computation of response functions and thus of electromagnetic cross sections.
In particular, the calculation of excited states in the continuum for
medium-mass nuclei constitutes an open problem.  At a given
 energy, the wave function is composed by many different
channels, corresponding to all possible partitions into
different fragments, which are difficult to calculate.

A method that allows to circumvent this issue is the Lorentz integral transform (LIT)~\cite{efros1994,Efl07}, which reformulates the problem in such a way that the explicit knowledge of all $|\Psi_n\ket$ in the continuum is not necessary.
The Lorentz integral transform is defined as
\begin{equation} \label{lorenzo}
  { L}(\omega_0,\Gamma )=\frac{\Gamma}{\pi}\int d\omega \frac{R(\omega)}{(\omega -\omega_0)
               ^2+\Gamma^2}\:\mbox{,}
\end{equation}   
where $\omega_0$ and $\Gamma$ are parameters, with $\Gamma > 0 ${.}  
By substituting $R(\omega)$ in Eq.~(\ref{lorenzo}) with the
expression from Eq.~(\ref{response}) and using the
completeness relation of the Hamiltonian eigenstates,
\begin{equation} \label{compl1}
   \sum_{ n} |{ \Psi_n} \rangle \langle { \Psi_n} | = 1 \:\mbox{,}
\end{equation}
one obtains
\begin{eqnarray}
&& 
\nonumber
 { L}(\omega_0,\Gamma)=\frac{\Gamma}{\pi}\,\times\\
\nonumber
&&\langle \Psi_0 | {\hat{\Theta}}^{\dagger}\frac{1}{\hat
              {H}-E_0-\omega_0+i\Gamma}\frac{1}{\hat{H}-E_0-\omega_0-i\Gamma}\hat{\Theta}|\Psi_0 \rangle \\
\label{lorenzog}
&& = \frac{\Gamma}{\pi}\bra \widetilde{\Psi} |\widetilde{\Psi}
  \rangle \:\mbox{.}
\end{eqnarray}
In this way, the LIT of the response function is basically proportional to the squared norm of the state
 $|\widetilde{\Psi}\ket$. This state is found as solution  of the Schr\"{o}dinger-like equation
\begin{equation} \label{psi1}
  (\hat{H}-z )|\widetilde{\Psi}\rangle ={\hat{\Theta}} | \Psi_0\rangle \,,
\end{equation}
where  $z=E_0+\omega_0+i\Gamma$,
for different values of the parameters $\omega_0$ and $\Gamma$.
Because of the fact that ${ L}(\omega_0,\Gamma)$ is finite, the unique solution $|\widetilde{\Psi} \rangle$ of
Eq.~(\ref{psi1}) has the same asymptotic boundary conditions as a
bound state.  Thus, only bound-state methods are required to solve this equation.

In the LIT approach one first computes
$ {L}(\omega_0,\Gamma)$  in a direct way, without
requiring the knowledge of $R(\omega)$. In a second step, 
the response function is obtained from a numerical inversion of 
$ {L}(\omega_0,\Gamma)$~\cite{efros1999,andreasi2005}.
The typical inversion procedure is based on a least-squares fit. First, we make an ansatz
for the shape of the response function, e.g., as
\begin{equation} 
\label{ansatz}
R(\omega)=\omega^{3/2} \exp \left( -\alpha \pi (Z-1) \sqrt{2\mu\over\omega}\right) 
\sum_i^N c_i e^{-\frac{\omega}{\beta i}}\,,
\end{equation}
where the exponential prefactor is a Gamow factor, assuming that the first channel
is the one proton knock out, so that the remaining nucleus has $(Z-1)$ protons. Here $\mu\sim
\frac{A-1}{A}m$ is the reduced mass with $m$ being the nucleon
mass. 
The least-squares fit is optimizing the coefficients $c_i$ so that the LIT of
Eq.~(\ref{ansatz}) is coinciding with the calculated one.
The coefficient $\beta$ is a non-linear fit parameter which is also varied in the fit procedure. Typically, one can change the
ansatz in Eq.~(\ref{ansatz}), vary $N$ and perform the inversion of LITs at different $\Gamma$
values to obtain an uncertainty of the inversion procedure.

For few-body systems, where a direct calculation of $|\Psi_n\ket$ even in the continuum case is possible,
it has been shown that the LIT method leads to an exact response function~\cite{lapiana2000,golak2002} with the full final state interaction included. Because the relevant equation to solve, Eq.~(\ref{psi1}),
is a bound-state equation, this method essentially circumvents the obstacle of the continuum calculation,
the price to pay being that a very good precision in the calculation of $L(\omega_0,\Gamma)$ is needed in order to obtain a stable inversion~\cite{Efl07}. 

The application of the LIT method used in conjunction with hyperspherical harmonics expansions to solve
Eq.~(\ref{psi1}) allowed, e.g., to perform studies of the photodisintegration of the six- and seven-body
nuclei~\cite{bacca2002,BaB04,BaA04}.

\subsection{Merging coupled-cluster theory with the Lorentz integral transform method}

Predictive ab initio calculations of electromagnetic reactions have
traditionally been limited to relatively light mass number, as
discussed above. Medium-mass and heavy nuclei are typically studied
with other theories, such as mean field based approaches and density
functional
theory~\cite{Erler2011,Nakatsukasa2012,Piekarewicz2012,RocaMaza_Paar},
which, despite being extremely useful, have a less direct connection
to quantum chromodynamics.

To surpass previous limitations of the ab initio approach, we have
merged the advantage of the LIT method of reducing the continuum
problem to the solution of a bound-state equation with the mild
computational scaling that characterizes coupled-cluster theory with
increasing mass number. This led to the introduction of a new
technique, which we call LIT-CC~\cite{Bacca:2013dma} and which
essentially is a coupled-cluster formulation of the LIT method.

Given that in coupled-cluster theory one introduces the exponential
ansatz and then one works with similarity transformation, in the
LIT-CC method Eq.~(\ref{psi1}) becomes simply
\begin{equation} \label{calc_rr}
 (\overline{H}_N-z) 
 |{\tilde{\Psi}}_R(z) \ket = \overline{\Theta}_N | \Psi_0^R \ket\;,
\end{equation}
where $ | \Psi_0^R \rangle \equiv |  \Phi_0 \rangle $ is the right ground state,
while $\overline{\Theta}_N$ is the similarity transformed normal-ordered electromagnetic operator
\begin{equation}\label{hbar}
 \overline{\Theta}_N = \exp(-\hat{T}) \hat{\Theta}_N \exp(\hat{T}) \,.
\end{equation}
The solution $| \tilde{\Psi}_R(z)\rangle $ of Eq.~(\ref{calc_rr}) is found as linear superposition of particle-hole excitations on top of the reference Slater determinant as
\begin{eqnarray} \label{eom1}\nonumber
|\widetilde{\Psi}_R(z) \rangle &=&\hat{{\cal R}}(z) | \Phi_0 \rangle =r_0+\hat{{\cal R}}_1 + \hat{{\cal R}}_2 \dots | \Phi_0 \rangle \;,
\end{eqnarray}
where 
\begin{eqnarray}
  \nonumber
  \hat{{\mathcal R}}(z)&=& r_0+ \sum_{i,a}r^a_i \hat{a}_a^\dag \hat{a}_i +\frac{1}{(2!)^2}\sum_{i,j,a,b}r^{ab}_{ij}\hat{a}_a^\dag\hat{a}^\dag_b\hat{a}_j \hat{a}_i  \\
  &+&\frac{1}{(3!)^2}\sum_{i,j,k,a,b,c}r^{abc}_{ijk}\hat{a}_a^\dag\hat{a}^\dag_b \hat{a}^{\dag}_c \hat{a}_k \hat{a}_j \hat{a}_i \\
  \nonumber
  & +&\frac{1}{(4!)^2}\sum_{i,j,k,l,a,b,c,d}r^{abcd}_{ijkl}\hat{a}_a^\dag\hat{a}^\dag_b \hat{a}^{\dag}_c \hat{a}^{\dag}_d \hat{a}_l \hat{a}_k \hat{a}_j \hat{a}_i  \\
& + & \dots \,.
  \end{eqnarray}
These operators build up  particle--hole ($p$--$h$) excitations and analogous expressions exist for the bra states.

If the expansions in the $\hat{T}$ and the $\hat{{\cal R}}$ operators are performed up to $Ap$--$Ah$, then the theory is exact. However, in practical applications one has to truncate these expansions.
The most common  approximation is the singles and doubles scheme D, for both the ground state and the excited states. In this work
we will investigate different approximation schemes in the ground
and excited states. In order to keep the notation concise we therefore
denote each scheme with a pair of labels (separated by a `$/$'
symbol), with the largest order of correlation included in the ground
state on the left, and the largest order of correlation included in
the excited states on the right. For example, when truncating
$\hat{T}=\hat{T}_1+\hat{T}_2$ and $\hat{{\mathcal R}}=r_0+\hat{{\mathcal R}}_1+\hat{{\mathcal R}}_2$, we will denote the calculation
with D/D.

\section{Interactions from $\chi$EFT}
\label{hamiltonians}

In the last decades we have observed the emergence, and its increased application, of
chiral effective field theories ($\chi$EFTs)
~\cite{Weinberg90,Epelbaum09,Machleidt11,Epelbaum12} to
systematically derive the interactions of nucleons among themselves and
with external electroweak probes. This approach allows to maintain a deeper connection to the underlying fundamental theory of quantum chromodynamics (QCD).
Effective Lagrangians,  expressed in terms of nucleons and pions,
are constructed so as  to preserve all symmetries, in particular the chiral symmetry,
characterizing QCD in the limit of vanishing quark masses.
The expansion
in powers of $(Q/\Lambda_\chi)^{\nu}$ is based on a separation of scales, where $Q$ is the low momentum characterizing  
low energy nuclear physics  and  $\Lambda_\chi\sim1$ GeV
is the chiral-symmetry breaking scale.
 The coefficients of the  expansion are called
Low-Energy Constants (LECs). They encapsulate the unresolved short-range physics and are adjusted to experimental data. This  
then makes it possible to predict nuclear observables to any degree $\nu$ of desired accuracy, with an
associated theoretical error roughly given by $(Q/\Lambda_\chi)^{(\nu+1)}$.  

In this approach, see, e.g.,
Refs.~\cite{Epelbaum09,Epelbaum12,Machleidt11}, three-nucleon (3N)
forces and higher-body forces arise naturally and consistently with
two-nucleon (NN) interactions. As such, they play an important role in
consistent calculations. In this paper, we will first show results
obtained only with NN forces~\cite{Entem03} and then also present
results using NN$+$3N forces. On the one hand we use a set of NN$+$3N
Hamiltonians starting from the next-to-next-to-next-to-leading order
(N$^3$LO) NN potential of Ref.~\cite{Entem03} evolved to lower
resolution scales using the similarity renormalization
group~\cite{bogner2007}. These low-momentum interactions are then
supplemented with a non-local 3N force at next-to-next-to-leading order
(N$^2$LO), adjusting the 3N LECs $c_D$ and $c_E$ to reproduce the
$^3$H binding energy and the $^4$He charge radius (for further details
see Ref.~\cite{Hebeler2011}). On the other hand we use the
NNLO$_{\rm sat}$ NN$+$3N Hamiltonian~\cite{Ekstroem2015} which was
adjusted to reproduce few-body observables as well as binding energies
and radii in selected nuclei up to mass number $A\approx25$. 
For this latter case, it is to be noted that  the NNLO$_{\rm sat}$ potential has been fit to experimental data using the $\Lambda$-CCSD(T) approximation  in a model space consisting of 15 oscillator shells and a frequency of 22 MeV. Here, we do not keep these model-space parameters fixed, but vary them in a truly ab initio spirit.  

The advantage of using a variety of interactions is that, by doing so, we
can give an estimate of systematic uncertainties of the employed
Hamiltonians. It also allows us to study whether different observables
are correlated, which in turn could be used to make predictions for
relevant quantities. Recent examples include the neutron radius and
dipole polarizability of $^{48}$Ca~\cite{Hagen2016}, the $2^+$ excited state in
$^{78}$Ni~\cite{hagen2016b}, and electromagnetic transitions in light
nuclei~\cite{calci2016}.

In all the calculations we will show, we use a model-space truncation
in $N_{\rm max}$, the number of harmonic-oscillator shells. While for
NN only we reach model-space sizes up to $N_{\rm
  max}=18$, the truncation for NN$+$3N calculations is $N_{\rm
  max}=14$, if not otherwise specified. Additionally, the 3N matrix elements are truncated in the
sum of the three-particle energies with typically  $E_{\rm 3max}\le16$.
The coupled-cluster computations start from a
Hartree-Fock reference state. For NN$+$3N calculations the 3N
contributions are included in normal-ordered two-body approximation,
discarding residual 3N forces. This approximation is valid for light
and medium-mass nuclei as shown in
Refs.~\cite{Hagen:2007ew,Roth:2011vt}.

\section{Results}
\label{results}

In this section we will present results obtained with the LIT-CC method. We will first focus on the dipole response function and photodisintegration cross section in Subsection~\ref{subsec:resp}. Then, we will show our studies of sum rules concentrating on the electric dipole polarizability $\alpha_D$ in Subsection~\ref{subsec:alphad}. Finally, we will address the role of triples corrections in Subsection~\ref{subsec:triples} and revisit the correlations among  $\alpha_D$ and nuclear radii for $^{48}$Ca in Subsection~\ref{subsec:48Ca}.

\subsection{Dipole response functions}
\label{subsec:resp}

The first electromagnetic reaction observables for which we exploited the power of coupled-cluster theory to address medium-mass nuclei has been the photodisintegration cross section.
In the unretarded dipole approximation valid at energies below the pion-production threshold, the photodisintegration cross section can be written as%
\begin{equation} 
\sigma_{\gamma}(\omega)=4\pi^2 \alpha \omega R(\omega)\,,
\label{cs}
\end{equation}
where  $\omega$ is the excitation energy and $R(\omega)$ is the dipole response function,
basically Eq.~(\ref{response}) where the electromagnetic operator is  the translationally invariant dipole 
\begin{equation}
\label{dip}
 {\hat{\Theta}}=\sum_k^A \left({\bf r}_k - {\bf R}_{\rm cm}  \right) \left( \frac{1+\tau^3_k}{2} \right)\,.
\end{equation}
Here ${\bf r}_k$ and ${\bf R}_{\rm cm}$ are the coordinates of the $k$-th particle and
the center-of-mass, respectively, while $(1+\tau^3_k)/2$  defines
the projection operator on the $Z$ protons, with $\tau^3_k$ being the third component of the $k$-th nucleon isospin. 

\begin{figure}[ht]
\centering
 \includegraphics*[width=9cm]{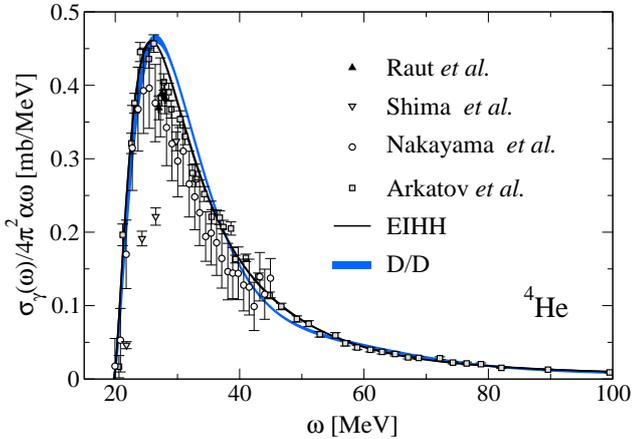}
\caption{\label{fig_4He} Dipole response function of the $^4$He nucleus. The LIT-CC calculation in the D/D approximation (thick curve) is compared to the exact one (thin curve) obtained from effective interaction hyperspherical harmonics (EIHH). A selected set of the available experimental data is shown for comparison. See text for details.}
\end{figure}

The photodisintegration cross section of Eq.~(\ref{cs}) has been calculated with the LIT method for a variety of light nuclei, from deuteron~\cite{lapiana2000} to $^7$Li~\cite{BaA04}, where the most extensive studies have been performed on $^4$He~\cite{PRL_4He}. In these cases the Schr\"{o}dinger-like equation was solved 
using exact few-body techniques.
Before the new LIT-CC method is used to study medium-mass nuclei, it is useful to benchmark it against 
the above mentioned exact calculations. $^4$He presents itself as an interesting case study, because it is 
a closed-shell nucleus, where the LIT-CC method can be most easily applied, and it has been substantially studied both from the theoretical
and the experimental point of view.
From the theory side, it is particularly instructive to test the D/D approximation against exact calculations performed with effective interaction hyperspherical harmonics~\cite{Barnea2000,Barnea2001}. In particular, we have performed such a comparison by using a two-body interaction derived in chiral effective field theory at N$^3$LO~\cite{Entem03}. 

Results of this comparison are shown in Fig.~\ref{fig_4He},
where we present the response function obtained after the inversion of the LIT.
In the inversion, we impose the response function to be zero before the threshold energy $\omega_{\rm
  th}$ which is the difference between the binding energy of
$^4$He and $^{3}$H, with $\gamma$ + $^4$He $\rightarrow$ $^3$H + $p$ being
the first open reaction channel in the photodisintegration process. 
Because, with the 
N$^3$LO potential the binding energies of $^4$He and $^3$H are not correctly
reproduced due to the missing 3N forces, a simple way to correct for that
is to shift the curves from the theoretical threshold $\omega_{\rm th}=17.54$ MeV  to the experimental one 
of 19.82 MeV. In this way, one can focus on the shape of the cross section in the comparison
to data. Three-nucleon forces were included, e.g., in Ref.~\cite{PRL_4He}, and their effect is 
to correct the threshold energy and decrease the peak height by a few percent.

The coupled-cluster D/D result is shown by the thick curve.
  The thickness of the curve is obtained 
  from inverting the LIT with
$\Gamma=10$~MeV and $\Gamma=20$~MeV and varying $N$ in Eq.~(\ref{ansatz}).  The EIHH results are instead represented
by the thin curve. In this case, by inverting LITs with $\Gamma=10$ and 20~MeV,
the two results overlap exactly.
Interestingly,  the D/D response function is close to the 
EIHH result, proving to be a very good approximation. Only small deviations  for energies between about
$\omega=30$ and $50$~MeV are seen. These are though much smaller than the uncertainties
of most of the experimental data available for the dipole response function.

In Fig.~\ref{fig_4He}, we show a selected set of data for comparison, see Ref.~\cite{Bacca:2014tla} for a more extensive discussion.
 The Arkatov {\it et al.}~\cite{Arkatov} data from the 70s  cover the broadest energy range.
They are in relatively good agreement with more recent  data by Raut {\it et al.}~\cite{Raut:2012zz,Tornow:2012zz},
where exclusive experiments were carried out measuring ($\gamma, p$) and ($\gamma, n$) at the High Intensity  Gamma-Ray source.
We summed up the two channels and show just the points in the energy range below the three-body break-up.
These first two sets of data obtained from  traditional photoabsorption experiments, agree rather well with dipole response function  data from Nakayama {\it et al.}~\cite{Nakayama} obtained via the study of the $^4$He($^7$Li,$^7$Be) reaction. A completely different trend is instead shown by the  Shima {\it et al.}~data from Ref.~\cite{Shima:2005ix},  obtained with a quasi-monoenergetic photon beam and a time projection chamber, where a simultaneous measurement of both the $^4$He$(\gamma,n)^3$He and  $^4$He$(\gamma,p)^3$H 
reactions has been performed.
Obviously the uncertainties in the experimental data (and the disagreement of about a factor of 2 of the Shima {\it et al.} data),
are much larger than the slight differences we observed between the EIHH and D/D coupled-cluster calculation. We impute this latter difference
to the missing triples and quadruple correlations and conclude that they can be safely neglected as their effect is small.

\begin{figure}
\centering
  \includegraphics[width=0.52\textwidth,clip=]{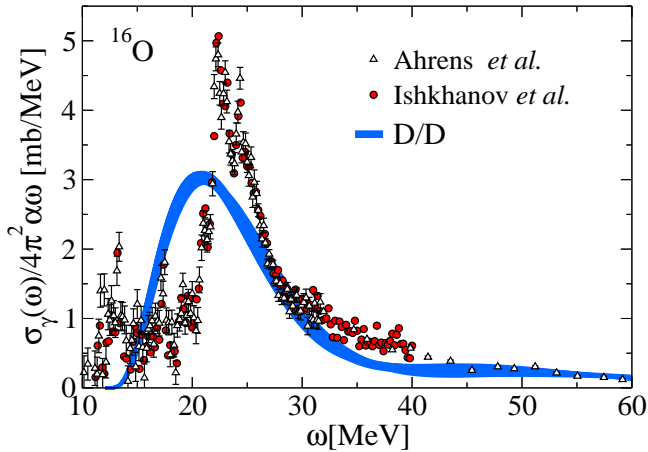}
  \caption{ $^{16}$O dipole response function in the D/D scheme calculated with the chiral N$^3$LO NN force~\cite{Entem03} and shifted to the experimental threshold,
   compared to data from Ahrens {\it et al.}~\cite{Ahrens75} and Ishkhanov {\it et al.}~\cite{ishkhanov2004}.}
  \label{fig_16O}
\end{figure}

After we  successfully benchmarked the new method   with exact hyperspherical harmonics on $^4$He, we  exploit the 
 mild computational scaling of the LIT-CC method with increasing mass number and investigate medium-mass nuclei, for the first time with an ab initio approach.

In Fig.~\ref{fig_16O}, we show the dipole response function of $^{16}$O calculated with the LIT-CC method at the D/D approximation level,
using  the same NN interaction  derived in chiral effective field theory at N$^3$LO~\cite{Entem03}. The curve is shown starting from the 
experimental threshold $\omega_{th}=12.1$ MeV.
The band thickness is obtained by inverting
the  LIT with 
width  $\Gamma=10$~MeV and by varying the number of basis
functions employed in the inversion.  Similar results are obtained by inverting the LIT at $\Gamma=20$~MeV.
We compare the theoretical results  with
experimental data by Ahrens {\it et al.}~\cite{Ahrens75}, who measured the total
photoabsorption cross section $\sigma_{\gamma}$ on an oxygen target
with natural abundance ($99.762\%$ $^{16}$O) with an attenuation
method. We also compare to a more recent evaluation by  Ishkhanov {\it et al.}~\cite{ishkhanov2004}.
We observe that the theoretical result is smeared compared to data, but overall the total dipole strength is correctly
reproduced, as well as  the bulk of the strength is in the right energy range. It is to note though that with this two-body interaction
and with the D/D approximation, we do not see the structures at around 10 MeV, which are found in experiment.

\begin{figure}
\centering
  \includegraphics[width=0.54\textwidth,clip=]{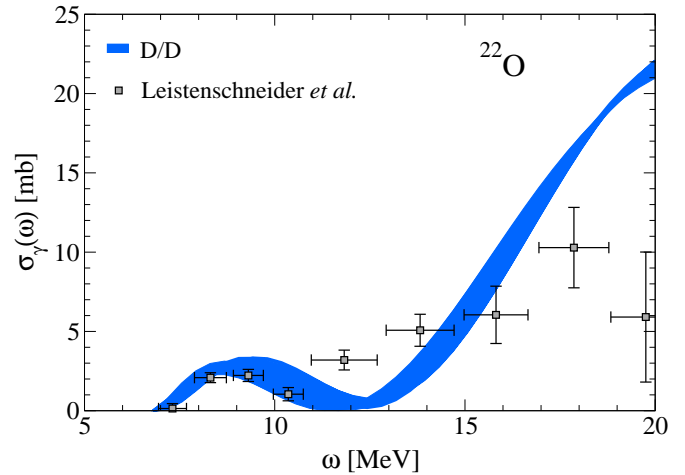}
  \caption{ $^{22}$O photodisintegration cross section calculated in the D/D scheme with the chiral N$^3$LO NN force~\cite{Entem03} and shifted to the experimental threshold, compared to data from Ref.~\cite{leistenschneider2001}.}
  \label{fig_22O}
\end{figure}

Another advantage of coupled-cluster theory is that one can also compute neutron-rich nuclei.
In particular, in the oxygen isotopic chain, a few of its neutron-rich isotopes have been studied
 at rare isotope beam facilities.
 The comparison of stable and unstable nuclei can provide key information about nuclear forces at the extremes of matter. Thus, it is
 interesting to apply these ab initio methods to exotic nuclei.
 We will focus on $^{22}$O, which is a closed sub-shell nucleus, for which the LIT-CC method 
can be applied. 

 In Fig.~\ref{fig_22O} we show the photoabsorption cross section of $^{22}$O computed in the D/D approximation~\cite{PRC2014} with the chiral N$^3$LO NN force~\cite{Entem03}. The width of the curve is obtained by performing several inversions,
for $\Gamma=20,10$ and 5 MeV.  We compare our results
 to experimental data taken at GSI by Leistenschneider~{\it et al.}~\cite{leistenschneider2001}. These are obtained from
a Coulomb excitation experiment and are turned into the equivalent of a photoabsorption cross section.
The experimental data show  a small peak at low energy. This structure is often named pygmy dipole resonance and was experimentally observed in  neutron-rich nuclei~\cite{Bracco2019}. Interestingly, our first principle calculation also presents a low-energy peak. The curve is shifted to start from the experimental threshold energy, as done for the other nuclei shown here. We remind the reader that the implemented two-body interaction  was tuned only on two-nucleon data. Despite that, we see the emergence of a pronounced  substructure at low energy as showed by the data.
At higher energies, the D/D results are larger than the data. This is expected because, while the experiment measured a semi-inclusive cross section the theoretical calculation is for an inclusive cross section, where  proton emission channels are included.

\begin{figure}[ht]
\centering
  \includegraphics[width=0.54\textwidth,clip=]{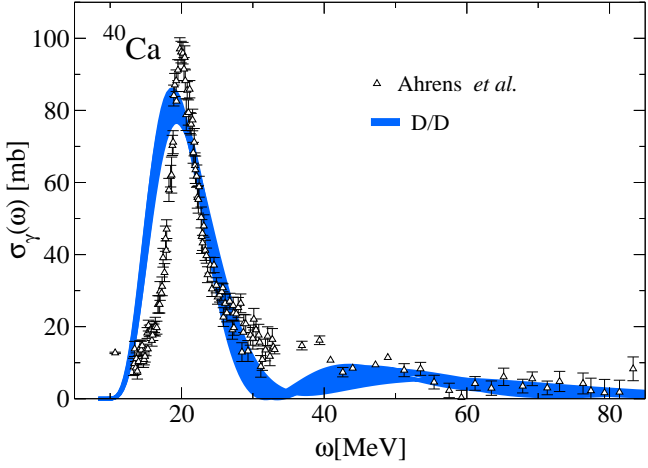}
  \caption{$^{40}$Ca photodisintegration
  cross section compared to  data from Ref.~\cite{Ahrens75}.  The curve is calculated with the chiral N$^3$LO NN force~\cite{Entem03} and is shifted to the  experimental threshold.}
  \label{fig_40Ca}
\end{figure}

The heaviest nucleus for which we calculated a response function with the LIT-CC method is  $^{40}$Ca.
In Fig.~\ref{fig_40Ca} we show the results in the D/D approximation obtained using the chiral N$^3$LO NN force~\cite{Entem03}
used for all other nuclei so far. The width of  the curve in Fig.~\ref{fig_40Ca} is obtained by inverting LITs with different $\Gamma$ parameters.
We compare it to the measured cross section by Ahrens {\it et al.}~\cite{Ahrens75}, where a natural calcium target was used with a photoabsorption attenuation method. The data
show  a very pronounced peak, referred to as the giant dipole resonance and located at an excitation energy of around  20 MeV.  This structure  is  well reproduced by the LIT-CC results.

As mentioned in the Introduction, first interpretations of such resonances,
 were given in terms of collective models~\cite{GoT48,steinwedel1950}. Now, with the advent of novel computational techniques such as the LIT-CC method we are able to show the emergence of these collective modes, both for the giant as well as for the pygmy resonance, from first principles.

Clearly, more work needs to be done to better assess the uncertainties included in the theoretical framework, as well as to include 3N forces. Nevertheless, the results shown in this Section constitute the first successful attempt to describe dipole response functions with ab initio methods. The main difficulty of this approach is that one needs to be able to compute the LIT very precisely, in order to be able to stably invert it. One limiting factor to achieve a sub-percentage convergence in the LIT calculation, is the availability of 
large model spaces, i.e., of computer memory and of matrix elements of the starting interaction. In all the above shown computations, the precision we could reach employing model spaces  of $N_{\rm max}=18$ was of the order of one to two percent, and that is the main reason why we obtain quite thick bands in the inversion procedure.

\subsection{Electric dipole polarizability}
\label{subsec:alphad}

If one is able to calculate the dipole response function, one can then easily compute its existing sum rules as
\begin{equation}
S_n=\int_{\omega_{th}}^{\infty} d\omega R(\omega) \omega^n \,,
\end{equation}
where $n$ is an integer number.
However, in several instances, it is easier to directly compute some selected sum rules than the whole response function, as the former may be computed as the expectation value of an operator on the ground state.

A particularly interesting sum rule of the dipole response function is the  electric dipole polarizability $\alpha_D$, which is
defined as
\begin{equation}\label{polresp}
\alpha_D = 2\alpha \int_{\omega_{th}}^{\infty}  d\omega~\frac{R(\omega)}{\omega}\,.
\end{equation}
It is basically an inverse energy weighted sum rule of the dipole response function.
As written in Eq.~(\ref{polresp}),
it is evident
that $\alpha_D$ contains the information on $R(\omega)$
at all energies $\omega$, including those in the continuum. A
calculation of $\alpha_D$ would then require to be able to solve the
many-body scattering problem in the continuum or to use alternative approaches as the LIT method. Indeed, for all the nuclei discussed in Subsection~\ref{subsec:resp}
it is easy to compute $\alpha_D$ by simply performing the integral in Eq.~(\ref{polresp}). 

On the other hand, using the completeness relations  it is possible to rewrite $\alpha_D$ as 
\begin{equation}
\label{sumrule} 
\alpha_D =\langle \Psi_0|\hat{\Theta}^\dag
\frac{1}{\hat{H}-E_0} \hat{\Theta} |\Psi_0\rangle\,.  
\end{equation}
So, if one is able to deal with the operator $\hat{\Theta}^\dag
\frac{1}{\hat{H}-E_0} \hat{\Theta}$, the calculation of $\alpha_D$ reduces to
an expectation value of the ground state, where no excited state is involved at all. However, that operator is not easy to write down, so practically what one does is to include
completeness of the  eigenstates (or a bound-state representation of them)
and use an eigenrepresentation of the Hamiltonian.
At this point, one can then see a relation between this equation with the second line of Eq.~(\ref{lorenzog}).
In fact, because in the limit $\Gamma \rightarrow 0$ 
the Lorentzian kernel becomes a delta function
\begin{equation}
\label{litdelta}
L(\omega_0,\Gamma\to 0) = \int R(\omega)\delta(\omega - \omega_0)d\omega=R(\omega_0)\,,
\end{equation}
$\alpha_D$ can be computed from the LIT as
\begin{equation}
\label{poldelta}
\alpha_D = 2\alpha\int \frac{L(\omega_0,\Gamma\to 0)}{\omega_0}d\omega_0\,.
\end{equation}
Equation~(\ref{litdelta}) is nothing else than a discretized response function and Eq.~(\ref{poldelta}) is just
an integral of that with an inverse energy weight. 
In Ref.~\cite{Miorelli2016} we showed that this method is equivalent to computing first the response function
with the LIT approach and then directly integrate the response function.
The advantage of using Eq.~(\ref{poldelta}) is that it does not require to invert $L(\omega_0,\Gamma)$ and have, consequently, an additional numerical error.
For this reason, if one is interested just in the electric dipole polarizability, it is preferable to use this method. In a slightly different spirit, one can understand that it is more complicated to compute the whole response function $R(\omega)$ because knowing the response functions means  knowing all of its existing sum rules, not just one.

\begin{table}[htb]
\centering
\caption{List of results with NN$+$3N Hamiltonians~\cite{Hebeler2011,Ekstroem2015} for the charge radius and the electric dipole polarizability for $^{16}$O and $^{40}$Ca in the D/D approximation. For the notation of the potentials we follow Ref.~\cite{Hebeler2011}.  Experimental values are taken from Ref.~\cite{angeli2013} (radius) and Ref.~\cite{Ahrens75,Miorelli2018,Birkhan17} (electric dipole polarizability). }
\label{table_data}
\begin{center}
\footnotesize
\renewcommand{\tabcolsep}{1.8mm}
\begin{tabular}{lll}
\hline
{$^{16}$O}&{}&{}\\
\hline
{Interaction}&{$R_{\rm ch}$ [fm]}& {$\alpha_D$ [fm$^3$]}\\
\hline
{2.0/2.0(EM)}&{2.62}&{0.46}\\
{2.0/2.0(PWA)}&{2.74}&{0.54}\\
{1.8/2.0(EM)}&{2.60}&{0.44}\\
{2.2/2.0(EM)}&{2.63}&{0.48}\\
{2.8/2.0(EM)}&{2.67}&{0.52}\\
{NNLO$_{\rm sat}$}&{2.71} & {0.58}\\
\hline
{Experiment}&{2.6991(52)~\cite{angeli2013}}& {0.58(1)~\cite{Ahrens75}}\\
{}&{ }& {0.568(9)~\cite{Miorelli2018}}\\
\hline
{$^{40}$Ca}&{}&{}\\
\hline
{2.0/2.0(EM)}&{3.35}&{1.67}\\
{2.0/2.0(PWA)}&{3.55}&{2.03}\\
{1.8/2.0(EM)}&{3.31}&{1.57}\\
{2.2/2.0(EM)}&{3.38}&{1.75}\\
{2.8/2.0(EM)}&{3.44}&{1.94}\\
{NNLO$_{\rm sat}$}&{3.48} & {2.08}\\
\hline
{Experiment}&{3.4776(19)~\cite{angeli2013}}& {2.23(3)~\cite{Ahrens75}}\\
{}&{}& {1.87(3)~\cite{Birkhan17}}\\
\hline
\end{tabular}
\end{center}
\end{table}

The  $L(\omega_0,\Gamma)$ in the LIT-CC approach is always computed using the Lanczos algorithm for non-symmetric matrices~\cite{PRC2014} and  in Ref.~\cite{Miorelli2016} we showed that $\alpha_D$ can be computed as continued fraction
 of the Lanczos coefficients~\cite{Lanczos50}.
 With this technology at hand, we can now explore the dependence of the
polarizability on the employed nuclear interaction.  For this purpose,
we will also use the potentials and normal-ordered 3N forces
introduced in Section~\ref{hamiltonians}.  For the notation of the
potentials we follow Ref.~\cite{Hebeler2011}.

In Table~\ref{table_data} we show results obtained within the D/D
approximation for the nuclear charge radius $R_{\rm ch}$ and for
$\alpha_D$ for $^{16}$O and $^{40}$Ca. Calculations were performed
with a model-space size of $N_{\rm max}=14$, for which we show the
central value obtained with the optimal harmonic-oscillator
frequency. The uncertainty associated with these D/D calculations were
estimated to be of about 1$\%$ for the radius and 2$\%$ for the
polarizability~\cite{Miorelli2016}.

One can readily see that, as expected, $\alpha_D$ and $R_{\rm ch}$ are
strongly correlated. At the bottom of the table we also show the
experimental results from Angeli and Marinova~\cite{angeli2013} for
the radius and from Ahrens {\it et al.}~\cite{Ahrens75} for the
polarizability. When integrating the experimental cross section
from  Ahrens {\it et al.}~\cite{Ahrens75}, a lower value for the polarizability
was obtained in Refs.~\cite{Miorelli2018,Birkhan17}.
 The agreement with data is particularly good for the
interaction NNLO$_{\rm{sat}}$, obviously so for the radius of
$^{16}$O, since it was fit to reproduce the experimental value, while
the polarizability and both observables for $^{40}$Ca are predictions
of this potential.  As shown in Ref.~\cite{Miorelli2016}, if one used
two-body forces only, one would still see a strong correlation, but
generally an underestimation of the experimental value for both the
charge radius and the polarizability.

We remind the reader that, while in Ref.~\cite{Miorelli2016} we were
able to compute the full $L(\omega_0,\Gamma)$ and invert it for
$^{4}$He and $^{16}$O with the NNLO$_{\rm sat}$ potential, the
inversion procedure was found to be much more unstable than for the
calculations with two-body forces only shown in
Subsection~\ref{subsec:resp}, primarily due to the fact that we are
limited in the model-space size to about $N_{\rm max}=14$ (as opposed
to $N_{\rm max}=18$ for two-body forces alone), which translate in a
larger uncertainty in the $L(\omega_0,\Gamma)$ and consequently on the
inversion.  Thus, from now on, we will continue our discussion on the
electric dipole polarizability only.

\subsection{Adding coupled-cluster triples corrections}
\label{subsec:triples}

Coupled-cluster theory is a systematically improvable method. So far
we showed results obtained within the coupled-cluster singles and
doubles approximation and have assessed uncertainties due to the
model-space expansion and residual dependence on the
harmonic-oscillator frequency, and by comparing to exact results in
light nuclei~\cite{Hagen2016,Ekstroem2015}. In order to improve the
results and more rigorously assess the uncertainty associated to the
coupled-cluster expansion, the next natural step is to include triples
corrections.  We include leading-order $3p$--$3h$ excitations using the
so called CCSDT-1 iterative triples approach~\cite{lee1984} which is a
good approximation to the full triples.  In the ground state, it
typically accounts for about $99\%$ of the correlation energy and
includes the leading-order contribution $\left(\hat{H}_N\hat{T}_2\right)_C$ (here
the index $C$ denotes connected terms~\cite{shavittbartlett2009}) to
the $\hat{T}_3$ amplitudes with an energy denominator given by the
Hartree-Fock single-particle energies, while all $\hat{T}_3$ contributions
to the $\hat{T}_1$ and $\hat{T}_2$ amplitudes are fully included. We  also
solve for the corresponding left ground state in the CCSDT-1 following
Ref.~\cite{watts1995}. Here, we denote this approximation with T-1,
consistently with Ref.~\cite{Miorelli2018}.  Analogously, for the equation of
motion and excited states we will add leading-order $3p$--$3h$
excitations in the T-1 approach~\cite{watts1995,jansen2016}. When T-1
is used both in the ground state and excited states, we will denote
this scheme as T-1/T-1.

When including $\hat{T}_3$ contribution  in the similarity transformation of a
normal-ordered one-body operator $\hat{\Theta}_N$  one has
\begin{eqnarray}
  \nonumber
  \overline{\Theta}_N&=&\left [ \hat{\Theta}_N \exp(\hat{T}_1+\hat{T}_2+\hat{T}_3)\right]_C= \\
\label{oddo1}
  &=& \overline{\Theta}^D_N + \left[ \hat{\Theta}_N \!\left(\!\frac{\hat{T}_2^2}{2}+\hat{T}_3+\hat{T}_1\hat{T}_3 \right) \right ]_C\,,
  \end{eqnarray}
where $\overline{\Theta}^D_N$ is the
similarity-transformed operator in the D approximation. 
In Ref.~\cite{Miorelli2018} we have shown that $ \overline{\Theta}_N$
can  safely be approximated with $\overline{\Theta}^D_N$, dramatically simplifying 
the calculations. Thus, we will only report on results obtained in this scheme.

At this point, it is interesting to compare calculations for the electric dipole polarizability with triples with those only including singles and doubles. Furthermore, because in the computation of $\alpha_D$ we need to specify an approximation scheme for the ground state and for the excited states, we also explore the case where T-1 triples are included only in the ground state, and denote this scheme with T-1/D. Obviously, this calculation is less computationally intensive than the T-1/T-1 case.

\begin{figure}
	\begin{center}
\includegraphics[width=\linewidth]{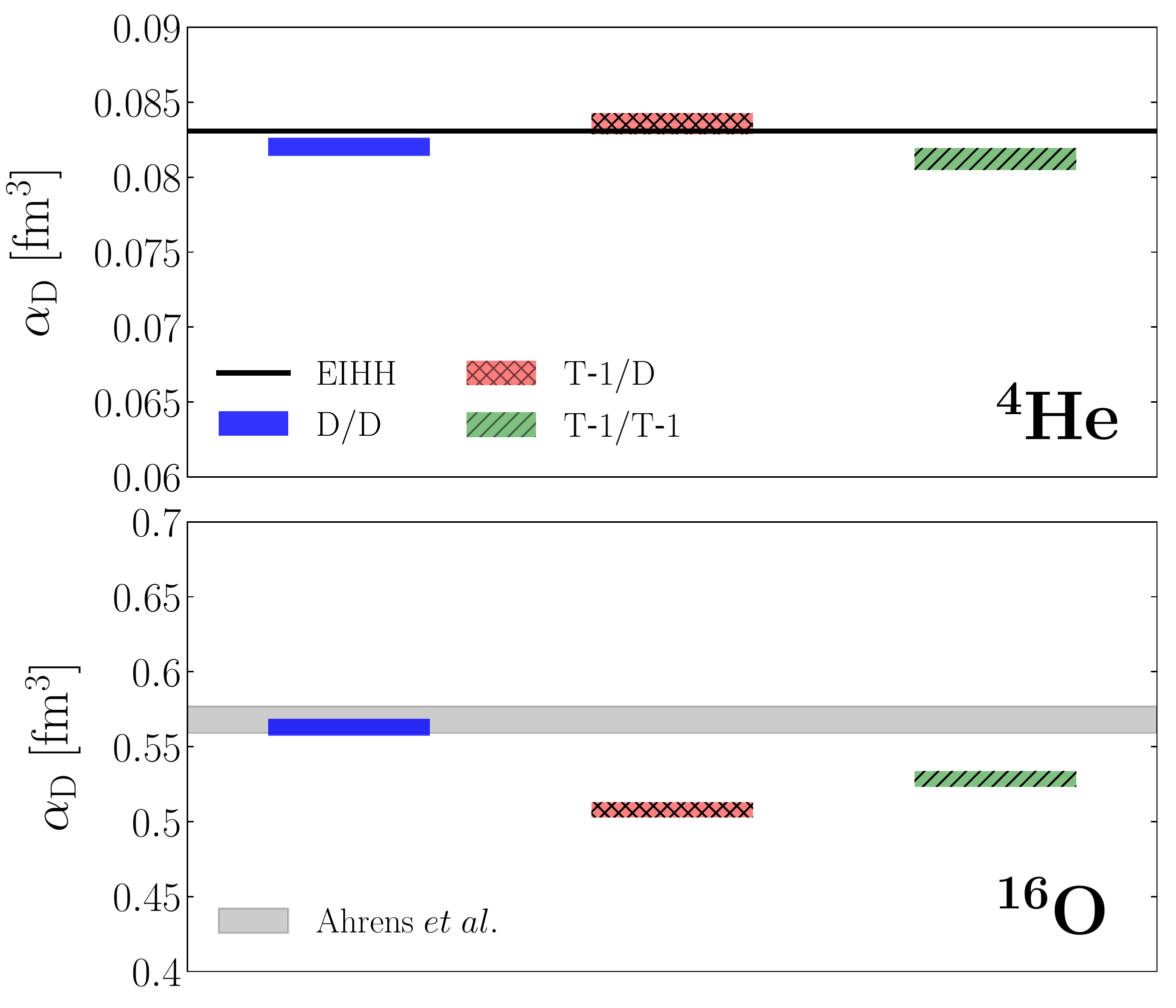}
\caption{Electric dipole polarizability for $^4$He (upper panel)
  and for $^{16}$O (lower panel) computed in the D/D (blue/left), the T-1/D (red/central) and the
  T-1/T-1 (green/right) approximations. For $^4$He the two-body force from Ref.~\cite{Entem03} is used and compared against the exact EIHH result. For $^{16}$O we use the NNLO$_{\rm sat}$ Hamiltonian~\cite{Ekstroem2015} with $N_{max}=12$  and
  compare to  what we obtain by integrating the experimental cross section from Ahrens {\it et al.}~\cite{Ahrens75}.}
      \label{fig:pol_triples}
  \end{center}
\end{figure}

In Fig.~\ref{fig:pol_triples}, we show $\alpha_D$ for $^4$He and $^{16}$O with the D/D, T-1/D and T-1/T-1 schemes.
We employ the chiral two-body force at N$^3$LO from \cite{Entem03} for $^4$He  because we want
to benchmark the various approximation schemes against virtually exact
EIHH, which cannot easily employ the
NNLO$_{\rm sat}$ Hamiltonian due to the non-locality of the 3N forces.
For $^{16}$O instead we use the NNLO$_{\rm sat}$ Hamiltonian~\cite{Ekstroem2015}, since our goal is to compare
with experimental data from Ahrens {\it et al.}~\cite{Ahrens75}.

For $^4$He we see that the T-1/D calculation agrees best with the hyperspherical
harmonics result shown by the black line, surprisingly better than the T-1/T-1 approach.
 The width of the bands reflects the residual
harmonic-oscillator dependence for the largest model space $N_{\rm
  max}=14$ and no cut has been done on the $3p$--$3h$ configurations.
Overall, the effect of
$3p$--$3h$ excitations in $^4$He is small, at the level of 1 $\%$, which is comparable to the
uncertainty obtained from the model-space variation and dependence on the harmonic-oscillator
frequency.

For $^{16}$O, while the D/D value is obtained at $N_{\rm max}=12$, 
when adding $3p$--$3h$ configurations we used a
model space of $N_{\rm max}=12$ and $E^F_{\rm 3max}=14$ for T-1/T-1
and T-1/D. Here, $E^F_{\rm 3max}$ is an energy cut on the allowed
$3p$--$3h$ excitations, defined as $|N_a-N_F|+|N_b-N_F|+|N_c-N_F|\le
E^F_{\rm 3max}$ and $|N_i-N_F|+|N_j-N_F|+|N_k-N_F|\le E^F_{\rm 3max}$
with a given harmonic-oscillator shell $N_p$ and the
harmonic-oscillator shell at the Fermi surface $N_F$.  The bands are
obtained by assigning a $2\%$ uncertainty, accounting for the combined
uncertainty from the $E^F_{\rm 3max}$ cut and the residual
harmonic-oscillator dependence.  Correlations arising from $3p$--$3h$
excitations reduce the size of $\alpha_D$, in this case both for
T-1/T-1 and T-1/D, by $8\%$ and $10\%$, respectively. Interestingly,
T-1/T-1 and T-1/D are very close to each other showing that including
triples corrections into the ground state is more important than
including them in the excited states.  The few percent difference
between the T-1/D and T-1/T-1 results can be taken as an estimate of
neglected higher-order correlations.  In comparison with the
experimental data from \cite{Ahrens75} shown by the grey band, we see
that the addition of triples leads to a further deviation of
$\alpha_D$ with respect to the experimental data, which agreed better
in the D/D approximation for this interaction. While the comparison to
experiment has to be seen as a judgment on the Hamiltonian itself, one
also has to keep in mind that the experimental value has been
extracted from sum rules from photoabsorption data that may be prone
to larger systematic uncertainties than those quoted, because it is
difficult to estimate the role of multipoles beyond the dipole.

\begin{table}[htb]
\centering
\caption{Effect of triples corrections for $\alpha_D$ of $^{16}$O in fm$^{3}$ using various NN$+$3N Hamiltonians~\cite{Hebeler2011,Ekstroem2015}.
  For the notation of the potentials we follow Ref.~\cite{Hebeler2011}.  The experimental value is taken from Refs.~\cite{Ahrens75,Miorelli2018}.}
\label{table_triples}
\begin{center}
\footnotesize
\renewcommand{\tabcolsep}{1.8mm}
\begin{tabular}{lll}
\hline
{$^{16}$O}&{}&{}\\
\hline
{Interaction}&{D/D}& {T-1/D}\\
\hline
{2.0/2.0(EM)}&{0.46}&{0.42}\\
{2.0/2.0(PWA)}&{0.54}&{0.48}\\
{1.8/2.0(EM)}&{0.44}&{0.41}\\
{2.2/2.0(EM)}&{0.48}&{0.43}\\
{2.8/2.0(EM)}&{0.52}&{0.45}\\
{NNLO$_{\rm sat}$} & {0.58}&{0.50}\\
\hline
{Experiment}&{}& {0.58(1)~\cite{Ahrens75}}\\
{}&{}& {0.568(9)~\cite{Miorelli2018}}\\
\hline
\end{tabular}
\end{center}
\end{table}

In Table~\ref{table_triples} we include new results for $^{16}$O
obtained within the T-1/D scheme for the electric dipole
polarizability, computed with all the Hamiltonians as in
Table~\ref{table_data}.  Interestingly, we find that with
triples corrections our results compare better with the value of 0.4959
fm$^{3}$ obtained by Raimondi and Barbieri~\cite{Raimondi} with the
self consistent Green's function method for the same NNLO$_{\rm
  sat}$ interaction. This might be coincidental as explained in
Ref.~\cite{Miorelli2018}, since coupled-cluster theory in the simple
D/S approximation also yields a similar value, namely 0.503 fm$^{3}$.
Overall, we find that the effect of coupled-cluster triples
corrections is lower for the softer interactions and higher for harder
interactions, reaching about 15$\%$ for the NNLO$_{\rm sat}$
case. For all the interactions, we get values that slightly
under-estimate the experimental data, which as mentioned before, could
potentially also include the effect of higher multipoles.

\subsection{Revisiting  correlations in $^{48}$Ca}
\label{subsec:48Ca}

\begin{figure*}
\centering
\includegraphics[width=17.5cm]{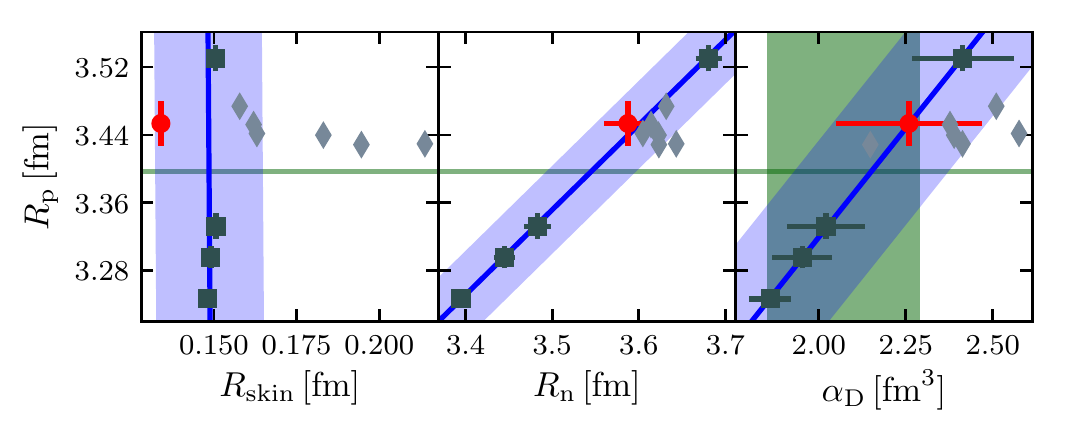}
\caption{ \label{fig_Ca48}  $^{48}$Ca  proton-distribution radius versus neutron-skin thickness (left panel), neutron-distribution radius (middle panel), and electric dipole polarizability (right panel). Results include leading triples excitations in coupled-cluster theory and are obtained by using various parameterization of the $\chi$EFT interactions~\cite{Hebeler2011} (squares) and the NNLO$_{\rm sat}$ (circle)~\cite{Ekstroem2015} potential. The correlation among these points are indicated by the (blue) light bands, while the experimental values for $R_p$~\cite{angeli2013} and $\alpha_D$~\cite{Birkhan17} are indicated with the (green) darker bands. 
 The diamonds correspond to the selected density functional theory calculations as also shown in Ref.~\cite{Hagen2016}.}
\end{figure*}

Recently, the first ab initio computation of the dipole
polarizability and the neutron-skin thickness of
$^{48}$Ca were provided~\cite{Hagen2016}.  $^{48}$Ca is a neutron-rich isotope of
calcium, with 20 protons and 28 neutrons. Despite the fact that it is
one of the candidates for neutrino-less double beta decay, its half
life is so long, that it can be considered as a stable nucleus and
studied in fixed target experiments.  In particular, an important
observable that is of great interest is the so-called neutron-skin
thickness, which is defined as the difference of the root mean square
distribution of neutrons, $R_n$, and of protons, $R_p$, in the nucleus
as
\begin{equation}
R_{\rm skin}= R_n -R_p \,.
\end{equation}
The neutron-skin thickness basically measures the extent of the
neutron distribution in a nucleus.  While nuclear charge (thus proton)
distributions are traditionally easily measured, e.g., with elastic
electron scattering, the distribution of neutrons is much more elusive
and difficult to access directly.  Presently, the cleanest way to
measure this quantity is expected to be parity violating electron scattering, which is mediated by the weak force through the exchange of a
$Z^0$ boson. Due to the fact that the proton-weak current is
approximately zero, this probe is very sensitive to the location of
the neutrons. There are plans to measure neutron skins
at Jefferson Laboratory with the CREX~\cite{CREX} and PREX experiments~\cite{PREX}, as well as in Mainz with the MREX n-skin project @MESA~\cite{MREX}.

\begin{table}[htb]
\centering
\caption{Effect of triples corrections in $^{48}$Ca for $R_{p}$, $R_{n}$, $R_{\rm skin}$ in fm and $\alpha_D$ in fm$^3$ using various NN$+$3N Hamiltonians~\cite{Hebeler2011,Ekstroem2015} as in Table~\ref{table_triples}. Calculations were performed for the harmonic-oscillator frequency of $\hbar\omega =16$ MeV.}
\label{table_triples_radii_ca48}
\begin{center}
\footnotesize
\renewcommand{\tabcolsep}{1.8mm}
\begin{tabular}{lll}
\hline
{$^{48}$Ca, $R_p$}&{}&{}\\
\hline
{Interaction}&{D}& {T-1}\\
\hline
{2.0/2.0(EM)}&{3.271}&{3.296}\\
{2.0/2.0(PWA)}&{3.502}&{3.530}\\
{1.8/2.0(EM)}&{3.225}&{3.247}\\
{2.2/2.0(EM)}&{3.307}&{3.332}\\
{NNLO$_{\rm sat}$} & {3.431}&{3.453}\\
\hline
\hline
{$^{48}$Ca, $R_n$}&{}&{}\\
\hline
{Interaction}&{D}& {T-1}\\
\hline
{2.0/2.0(EM)}&{3.423}&{3.445}\\
{2.0/2.0(PWA)}&{3.656}&{3.681}\\
{1.8/2.0(EM)}&{3.375}&{3.395}\\
{2.2/2.0(EM)}&{3.461}&{3.483}\\
{NNLO$_{\rm sat}$} & {3.571}&{3.587}\\
\hline
{$^{48}$Ca, $R_{\rm skin}$}&{}&{}\\
\hline
{Interaction}&{D}& {T-1}\\
\hline
{2.0/2.0(EM)}&{0.151}&{0.149}\\
{2.0/2.0(PWA)}&{0.154}&{0.150}\\
{1.8/2.0(EM)}&{0.150}&{0.148}\\
{2.2/2.0(EM)}&{0.154}&{0.151}\\
{NNLO$_{\rm sat}$} & {0.139}&{0.134}\\
\hline
{$^{48}$Ca, $\alpha_D$}&{}&{}\\
\hline
{Interaction}&{D/D}& {T-1/D}\\
\hline
{2.0/2.0(EM)}&{2.118}&{1.952}\\
{2.0/2.0(PWA)}&{2.693}&{2.414}\\
{1.8/2.0(EM)}&{1.979}&{1.860}\\
{2.2/2.0(EM)}&{2.233}&{2.021}\\
{NNLO$_{\rm sat}$} & {2.645}&{2.259}\\
\hline
\end{tabular}
\end{center}
\end{table}

In Ref.~\cite{Hagen2016} a study of the correlations of $\alpha_D$
with $R_n$ and $R_{\rm skin}$ was presented, that provided theoretical predictions
for all the three above mentioned quantities targeted by
modern experiments.  The electric dipole polarizability of $^{48}$Ca
was indeed recently measured with $(p,p')$ reactions in Osaka and was found to be in 
rather good agreement with coupled-cluster calculations~\cite{Birkhan17},
albeit somewhat smaller.

The radii and polarizability calculations in Refs.~\cite{Hagen2016}
were performed in the D/D scheme. Here, we
revisit our results using the recently developed $3p$--$3h$
technology~\cite{Miorelli2018}. In Ref.~\cite{Miorelli2018} we already
computed $\alpha_D$ for $^{48}$Ca in the T-1/D scheme for a couple of
interactions and found that its values is reduced, improving the
agreement with the experimental measurement by Birkhan {\it et
  al.}~\cite{Birkhan17}.  Here, we supplement our results with a few
more Hamiltonians and present complete results for radii and the dipole polarizability
 in Table \ref{table_triples_radii_ca48} (with an increased number of decimal digits compared to the other tables in this work for clarity of the following discussion). 
On the one hand, we find that triples corrections only mildly affect the calculations
of $R_p$ and $R_n$, leading to an increase in both quantities of less than $1\%$. Triples effects are slightly 
larger on $R_p$ than on $R_n$, and as a consequence $R_{\rm skin}$ calculated in the T-1/D scheme is 
smaller than $R_{\rm skin}$ calculated in the D/D approximation. 
On the other hand, triples correlations affect $\alpha_D$ quite visibly, as already pointed out in Ref.~\cite{Miorelli2018}. In particular, for the harder interactions, their effect is of about $15\%$.

In Fig.~\ref{fig_Ca48} we plot these results, showing the proton radius as a function of the skin radius, the
neutron-distribution radius and the electric dipole polarizability.
 The  squares correspond to coupled-cluster calculations in the T-1/D scheme
using four different parameterization of the $\chi$EFT
interactions~\cite{Hebeler2011}, while the  circles correspond to the
results obtained with the
NNLO$_{\rm sat}$~\cite{Ekstroem2015} potential. 
Each theory point is plotted with
error bars that include both the residual $\hbar \omega$-dependence, as well as
an estimate of the coupled-cluster truncation error. Below we briefly explain the recipe we use to estimate uncertainties.

The uncertainty $\delta_\CO$ for any observable $\CO$ ($R_p, R_n, R_{\rm skin}$ and $\alpha_D$)  is computed as the quadrature of
\begin{equation}
\delta_\CO=\sqrt{(\delta^{\hbar \omega}_\CO)^2+(\delta^{\text{CC}}_\CO)^2}\,,
\end{equation}
where $\delta^{\hbar \omega}_\CO$ is an estimate of the uncertainty related to the residual harmonic oscillator $\hbar \omega$-dependence, while $\delta^{\text{CC}}_\CO$ is
an estimate of the coupled-cluster truncation uncertainty. For the latter, in case of the D/D calculations we take the difference from the T-1/D and D/D results, while in case of the T-1/D calculations, given that we do not have any higher order coupled-cluster calculation available, we take half of the above mentioned difference. 
More formally stated, the two uncertainties summed in quadrature are taken to be
\begin{align}
\nonumber
\delta^{\hbar\omega}_\CO&=\frac{\CO(\hbar\omega_1)-\CO(\hbar\omega_2)}{2}\,,\\
\delta^{\rm CC}_\CO&=\CO^{\text{D/D}}(\hbar\omega_1)-\CO^{\text{T-1/D}}(\hbar\omega_1)
\end{align}
for the D/D calculations, and
\begin{align}
\nonumber
\delta^{\hbar\omega}_\CO&=\frac{\CO(\hbar\omega_1)-\CO(\hbar\omega_2)}{2}\,,\\
\delta^{\rm CC}_\CO&=0.5\cdot\left ( \CO^{\text{D/D}}(\hbar\omega_1)-\CO^{\text{T-1/D}}(\hbar\omega_1) \right)
\end{align}
for the T-1/D calculations.
The values are computed for the maximum available model space size of  $N_{\mathrm max}=14$, using two neighboring frequencies around the optimal value, namely $\hbar \omega_1=16$ MeV and $\hbar \omega_2=12$ MeV.
The uncertainties of $R_{\rm skin}$ are obtained taking the correlation between $R_p$ and $R_n$
into account using standard covariance theory.
Clearly, this is a rough uncertainty estimate, but we find it sensible for the purpose of updating Ref.~\cite{Hagen2016}.

From Fig.~\ref{fig_Ca48}, it is clear that the strong correlation between $R_p$, $R_n$ and $\alpha_D$, which was observed in \cite{Hagen2016}, is confirmed, as well as we reinforce the fact that $R_{\rm skin}$ is almost constant with respect to the employed interaction.
As done in Ref.~\cite{Hagen2016}, one can exploit the
correlation among observables with the fact that the proton
distribution radius is known experimentally,  and from the intersection between the correlation band and the horizontal line, one can draw
constraints on the neutron-skin thickness and the
electric dipole polarizability. In Fig.~\ref{fig_Ca48} we do not explicitly show these constraints, but rather highlight the linear correlation bands and the experimental constraints with their corresponding uncertainty.
 With respect to Ref.~\cite{Hagen2016}, revisiting the calculations with leading triples, we obtain
that  the constraints on the dipole polarizability move from  $2.19 \le \alpha_D \le 2.60$ fm$^3$ to
 $1.92 \le \alpha_D \le 2.38$ fm$^3$, thus getting closer to the experimental value of 2.07(22) fm$^3$~\cite{Birkhan17}, while the constraints for the neutron-skin thickness go from  $0.12 \le R_{\rm skin}
\le 0.15$ fm to $0.13 \le  R_{\rm skin} \le 0.16$ fm, thus not varying much.
Despite the fact that $R_{\rm skin}$ becomes smaller with triples for a given model space parameter set, the width of the neutron-skin thickness constraint, as well as the central value of $R_{\rm skin}$ are slightly increased 
with respect to \cite{Hagen2016}. This is due to a combination of facts: the use of symmetric correlation bands; the use of five instead of six~\cite{Hagen2016} different interactions, and  the use of optimal harmonic-oscillator frequencies, where the convergence in $N_{max}$ is faster.
The difference is anyway small and the updated $^{48}$Ca study further establishes that the ab
initio  prediction  of the neutron-skin thickness is much smaller than the values obtained from selected density functional theory calculations, shown in  Fig.~\ref{fig_Ca48} by the diamonds.
Thus,  it is worth to reiterate that it is  important to have a clean, as model-independent as possible,  experimental determination of $R_{\rm skin}$.

\section{Conclusions}
\label{conclusions}
In this paper we review the recent progress made in the computation of
electromagnetic response functions and related sum rules using a
coupled-cluster theory formulation of the Lorentz integral transform
method. We present photoabsorption cross sections of $^{16, 22}$O and
$^{40}$Ca, showing that we obtain a reasonable description of the
experimental data already with two-body forces. We also review our
calculations of the electric dipole polarizability, for which we use
NN$+$3N Hamiltonians.  We also present new results for the
polarizability of $^{16}$O and $^{48}$Ca that include leading triples
coupled-cluster correlations. For the $^{48}$Ca case, we 
revisit our  previous studies of correlations among the electric dipole polarizability, the neutron-, proton- and  skin-radius. We show that  correlations still hold and allow to draw  improved predictions for the value of the polarizability, which is in better agreement with experiment, as well as for the neutron-skin thickness, for which we corroborate the earlier finding that ab initio theory predicts a  smaller neutron skin than density functional theory.\\

 {\it Acknowledgments.}-- This work was supported by the Deutsche
 Forschungsgemeinschaft (DFG) through the Collaborative Research
 Center [The Low-Energy Frontier of the Standard Model (SFB 1044)]; by
 the Cluster of Excellence “Precision Physics, Fundamental
 Interactions, and Structure of Matter” (PRISMA$^+$ EXC 2118/1) funded
 by DFG within the German Excellence Strategy (Project ID 39083149)
 and by the Office of Nuclear Physics, U.S. Department of Energy,
 under grants desc0018223 (NUCLEI SciDAC-4 collaboration) and by the
 Field Work Proposal ERKBP72 at Oak Ridge National Laboratory (ORNL).
 Computer time was provided by the Innovative and Novel Computational
 Impact on Theory and Experiment (INCITE) program. The new
 calculations presented in this work were also performed on ``Mogon
 II'' at Johannes Gutenberg-Universit\"{a}t in Mainz.
%

\end{document}